\documentstyle[amssymb,aps]{revtex}

\begin{document}
\title{{\it Brief Report}\\
A simple scheme of teleportation of arbitrary multipartite qubit
entanglement }
\author{Zai-Zhe Zhong}
\address{Department of Physics, Liaoning Normal University, Dalian 116029, Liaoning,\\
China. E-mail: zhongzaizheh@hotmail.com}
\maketitle

\begin{abstract}
In this paper, we define a cross product operator and construct the cross
Bell basis, by use this basis and Bell measurements we give a simple scheme
of the teleportation of arbitrary multipartite qubit entanglement.

PACC numbers: 03.67.Mn, 03.65.Ud, 03.67.Hk.
\end{abstract}

In modern quantum mechanics and quantum information, the quantum
teleportation is a quite interesting and important topic. Following the
BBCJPW scheme[1], there have been very many related works (e.g. see the
references in [2]). Generally, these works discussed the teleportaion of the
single unknown qubit states. Recently, Rigolin[3] gives a schemes of the
teleportation of arbitrary multipartite qubit states, there are yet some
relational improvements (e.g., see [4]). However these schemes are more
complex whatever in physics and in mathematical forms. Generally, in a
perfect quantum teleportation scheme, the most basic matter are to find the
quantum channels (they best form a basis of the a Hilbert space) and to give
a real physical measurement way distinguishing the outcomes of the wave
function collapses. In this paper, we give a simple scheme of teleportation
of arbitrary multipartite qubit entanglement. Our ways are to define a
convenient operator `cross products', to construct the `cross Bell basis'
and to use two or more common Bell measurements.

In the following we write the Hilbert space of states of a spin-$\frac 12$
particle $x$ as $H_x$, in which a pure-states $\mid \Psi _x\rangle
=\sum\limits_{i=0,1}c_i\mid i_x\rangle .$ In this paper, we mainly discuss
the Hilbert space $H_1\otimes H_2\otimes H_3\otimes H_4.$

{\it Definition. }Suppose that $\mid \Psi _{13}\rangle
=\sum\limits_{i,j=0,1}c_{ij}\mid i_1\rangle \mid j_3\rangle \in H_1\otimes
H_3,$ $\mid \Phi _{24}\rangle =\sum\limits_{r,s=0,1}d_{rs}\mid r_2\rangle
\mid s_4\rangle \in H_2\otimes H_4$ are two pure-states$,$ then the cross
product $\mid \Psi _{13}\rangle \triangledown \mid \Phi _{24}\rangle \in
H_1\otimes H_2\otimes H_3\otimes H_4$ of $\mid \Psi _{13}\rangle $ and $\mid
\Phi _{24}\rangle $ is defined to be the result of $\mid \Psi _{13}\rangle
\otimes \mid \Phi _{24}\rangle $ returning to the natural order 1,2,3,4,
i.e. 
\begin{equation}
\mid \Psi _{13}\rangle \triangledown \mid \Phi _{24}\rangle
=\sum_{i,r,j,s=0,1}c_{ij}d_{rs}\mid i_1\rangle \mid r_2\rangle \mid
j_3\rangle \mid s_4\rangle
\end{equation}

Notice that since the order of $H_1,H_2,H_3,H_4$ is important in our
discussion, so the operations $\otimes $ and $\triangledown $ are distinct.
In addition, obviously $\triangledown $ is a bi-linear and non-commutative
operator.

Now, we read the ordinary Bell bases as 
\begin{equation}
\mid \Psi _{\alpha \beta }^{\pm }\rangle =\frac 1{\sqrt{2}}\left( \mid
0_\alpha \rangle \mid 0_\beta \rangle \pm \mid 1_\alpha \rangle \mid 1_\beta
\rangle \right) ,\mid \Phi _{\alpha \beta }^{\pm }\rangle =\frac 1{\sqrt{2}}%
\left( \mid 0_\alpha \rangle \mid 1_\beta \rangle \pm \mid 1_\alpha \rangle
\mid 0_\beta \rangle \right)
\end{equation}
then by using crossed products and according to the rule similar to matrix
entries, we can write a group ${\Bbb B=}\left\{ K\triangledown L\right\} $
of sixteen pure-states $K\triangledown L$ as 
\begin{equation}
{\Bbb B=} 
\begin{array}{lllll}
^{_{K\triangledown L}} & ^{_{_{\Psi _{24}^{+}}}} & ^{_{_{\Psi _{24}^{-}}}} & 
^{_{^{_{\Phi _{24}^{+}}}}} & ^{_{_{\Phi _{24}^{-}}}} \\ 
^{_{\Psi _{13}^{+}}} & \mid \Psi _{13}^{+}\rangle \triangledown \mid \Psi
_{24}^{+}\rangle , & \mid \Psi _{13}^{+}\rangle \triangledown \mid \Psi
_{24}^{-}\rangle , & \mid \Psi _{13}^{+}\rangle \triangledown \mid \Phi
_{24}^{+}\rangle , & \mid \Psi _{13}^{+}\rangle \triangledown \mid \Phi
_{24}^{-}\rangle \\ 
^{_{\Psi _{13}^{-}}} & \mid \Psi _{13}^{-}\rangle \triangledown \mid \Psi
_{24}^{+}\rangle , & \mid \Psi _{13}^{-}\rangle \triangledown \mid \Psi
_{24}^{-}\rangle , & \mid \Psi _{13}^{-}\rangle \triangledown \mid \Phi
_{24}^{+}\rangle , & \mid \Psi _{13}^{-}\rangle \triangledown \mid \Phi
_{24}^{-}\rangle \\ 
^{_{\Phi _{13}^{+}}} & \mid \Phi _{13}^{+}\rangle \triangledown \mid \Psi
_{24}^{+}\rangle , & \mid \Phi _{13}^{+}\rangle \triangledown \mid \Psi
_{24}^{-}\rangle , & \mid \Phi _{13}^{+}\rangle \triangledown \mid \Phi
_{24}^{+}\rangle , & \mid \Phi _{13}^{+}\rangle \triangledown \mid \Phi
_{24}^{-}\rangle \\ 
^{_{\Phi _{13}^{-}}} & \mid \Phi _{13}^{-}\rangle \triangledown \mid \Psi
_{24}^{+}\rangle , & \mid \Phi _{13}^{-}\rangle \triangledown \mid \Psi
_{24}^{-}\rangle , & \mid \Phi _{13}^{-}\rangle \triangledown \mid \Phi
_{24}^{+}\rangle , & \mid \Phi _{13}^{-}\rangle \triangledown \mid \Phi
_{24}^{-}\rangle
\end{array}
\end{equation}
It is easily verified that ${\Bbb B}$ is a complete orthogonal basis of $%
H_1\otimes H_2\otimes H_3\otimes H_4$, we call it the `crossed Bell basis'.
Here it must be stressed that these bases are really distinguishable by Bell
measurements. For instance, for any state $\mid \Psi _{1234}\rangle \in
H_1\otimes H_2\otimes H_3\otimes H_4$ if we make two independent Bell
measurements jointed particle pair $\left( 1,3\right) $ and jointed particle
pair $\left( 2,4\right) $ respectively, then $\mid \Psi _{1234}\rangle $
must collapse to one of the above sixteen crossed Bell bases with a
probability. In the following, we notice $\stackrel{\vee }{x}=1-x$ for $x=0$
or $1.$ The transformation from the natural basis to the crossed Bell basis
is 
\begin{eqnarray}
&\mid &i_1\rangle \mid r_2\rangle \mid i_3\rangle \mid r_4\rangle =\frac 12%
\left( \mid \Psi _{13}^{+}\rangle +\mid \Psi _{13}^{-}\rangle \right)
\triangledown \left( \mid \Psi _{24}^{+}\rangle +\mid \Psi _{24}^{-}\rangle
\right)  \nonumber \\
&\mid &i_1\rangle \mid r_2\rangle \mid i_3\rangle \mid \stackrel{\vee }{r}%
_4\rangle =\frac 12\left( -1\right) ^r\left( \mid \Psi _{13}^{+}\rangle
+\mid \Psi _{13}^{-}\rangle \right) \triangledown \left( \mid \Phi
_{24}^{+}\rangle +\mid \Phi _{24}^{-}\rangle \right)  \nonumber \\
&\mid &i_1\rangle \mid r_2\rangle \mid \stackrel{\vee }{i}_3\rangle \mid
r_4\rangle =\frac 12\left( -1\right) ^i\left( \mid \Phi _{13}^{+}\rangle
+\mid \Phi _{13}^{-}\rangle \right) \triangledown \left( \mid \Psi
_{24}^{+}\rangle +\mid \Psi _{24}^{-}\rangle \right) \\
&\mid &i_1\rangle \mid r_2\rangle \mid \stackrel{\vee }{i}_3\rangle \mid 
\stackrel{\vee }{r}_4\rangle =\frac 12\left( -1\right) ^{i+r}\left( \mid
\Phi _{13}^{+}\rangle +\mid \Phi _{13}^{-}\rangle \right) \triangledown
\left( \mid \Phi _{24}^{+}\rangle +\mid \Phi _{24}^{-}\rangle \right) 
\nonumber
\end{eqnarray}

Now we prove that by using any one of cross Bell bases, we can realize the
teleportation of a unknown bipartite qubit pure-state. For instance, we take 
$\mid \Phi _{13}^{+}\rangle $ $\triangledown \mid \Phi _{24}^{-}\rangle $,
as the quantum channel, particles 3, 4 are in Alice. The receiptor is Bob,
she is in remote place from Alice, and she holds particles 1, 2. Suppose
that $\mid \varphi _{56}\rangle =\alpha \mid 0_50_6\rangle +\beta \mid
0_51_6\rangle +\gamma \mid 1_50_6\rangle +\delta \mid 1_51_6\rangle $ is a
client unknown state in Alice. It is known[5] that $\mid \psi _{56}\rangle $
is entangled if and only if $\alpha \gamma -\beta \delta $ $\neq 0$.

In the present case the total state is 
\begin{eqnarray}
&\mid &\Psi _{123456}\rangle =\left( \mid \Phi _{13}^{+}\rangle
\triangledown \mid \Phi _{24}^{-}\rangle \right) \otimes \mid \varphi
_{56}\rangle  \nonumber \\
&=&\frac 12\left( \mid 0_10_21_31_4\rangle -\mid 0_11_21_30_4\rangle +\mid
1_10_20_31_4\rangle -\mid 1_11_20_30_4\rangle \right) \\
&&\otimes \left( \alpha \mid 0_50_6\rangle +\beta \mid 0_51_6\rangle +\gamma
\mid 1_50_6\rangle +\delta \mid 1_51_6\rangle \right)  \nonumber
\end{eqnarray}
Expanding $\mid \Psi _{123456}\rangle $ and by using the formulae in Eq.(2)
to particles 3, 4, 5 and 6, the final result is 
\begin{equation}
\mid \Psi _{123456}\rangle =\left\{ 
\begin{array}{c}
\frac 14\left( \delta \mid 0_10_2\rangle +\gamma \mid 0_11_2\rangle -\beta
\mid 1_10_2\rangle -\alpha \mid 1_11_2\rangle \right) \Psi
_{35}^{+}\triangledown \Psi _{46}^{+} \\ 
+\frac 14\left( \delta \mid 0_10_2\rangle +\gamma \mid 0_11_2\rangle +\beta
\mid 1_10_2\rangle -\alpha \mid 1_11_2\rangle \right) \Psi
_{35}^{+}\triangledown \Psi _{46}^{-} \\ 
+\frac 14\left( \gamma \mid 0_10_2\rangle +\delta \mid 0_11_2\rangle -\alpha
\mid 1_10_2\rangle -\beta \mid 1_11_2\rangle \right) \Psi
_{35}^{+}\triangledown \Phi _{46}^{+} \\ 
+\frac 14\left( -\gamma \mid 0_10_2\rangle +\delta \mid 0_11_2\rangle
+\alpha \mid 1_10_2\rangle -\beta \mid 1_11_2\rangle \right) \Psi
_{35}^{+}\triangledown \Phi _{46}^{-} \\ 
\  \\ 
+\frac 14\left( -\delta \mid 0_10_2\rangle -\gamma \mid 0_11_2\rangle -\beta
\mid 1_10_2\rangle -\alpha \mid 1_11_2\rangle \right) \Psi
_{35}^{-}\triangledown \Psi _{46}^{+} \\ 
+\frac 14\left( \delta \mid 0_10_2\rangle -\gamma \mid 0_11_2\rangle +\beta
\mid 1_10_2\rangle -\alpha \mid 1_11_2\rangle \right) \Psi
_{35}^{-}\triangledown \Psi _{46}^{-} \\ 
+\frac 14\left( -\gamma \mid 0_10_2\rangle -\delta \mid 0_11_2\rangle
-\alpha \mid 1_10_2\rangle -\beta \mid 1_11_2\rangle \right) \Psi
_{35}^{-}\triangledown \Phi _{46}^{+} \\ 
+\frac 14\left( \gamma \mid 0_10_2\rangle -\delta \mid 0_11_2\rangle +\alpha
\mid 1_10_2\rangle -\beta \mid 1_11_2\rangle \right) \Psi
_{35}^{-}\triangledown \Phi _{46}^{+} \\ 
\  \\ 
+\frac 14\left( \beta \mid 0_10_2\rangle +\alpha \mid 0_11_2\rangle -\delta
\mid 1_10_2\rangle -\gamma \mid 1_11_2\rangle \right) \Phi
_{35}^{+}\triangledown \Psi _{46}^{+} \\ 
+\frac 14\left( -\beta \mid 0_10_2\rangle +\alpha \mid 0_11_2\rangle +\delta
\mid 1_10_2\rangle -\gamma \mid 1_11_2\rangle \right) \Phi
_{35}^{+}\triangledown \Psi _{46}^{-} \\ 
+\frac 14\left( \alpha \mid 0_10_2\rangle +\beta \mid 0_11_2\rangle -\gamma
\mid 1_10_2\rangle -\delta \mid 1_11_2\rangle \right) \Phi
_{35}^{+}\triangledown \Phi _{46}^{+} \\ 
+\frac 14\left( -\alpha \mid 0_10_2\rangle +\beta \mid 0_11_2\rangle +\gamma
\mid 1_10_2\rangle -\delta \mid 1_11_2\rangle \right) \Phi
_{35}^{+}\triangledown \Phi _{46}^{-} \\ 
\  \\ 
+\frac 14\left( -\beta \mid 0_10_2\rangle -\alpha \mid 0_11_2\rangle -\delta
\mid 1_10_2\rangle -\gamma \mid 1_11_2\rangle \right) \Phi
_{35}^{-}\triangledown \Psi _{46}^{+} \\ 
+\frac 14\left( \beta \mid 0_10_2\rangle -\alpha \mid 0_11_2\rangle +\delta
\mid 1_10_2\rangle -\gamma \mid 1_11_2\rangle \right) \Phi
_{35}^{-}\triangledown \Psi _{46}^{-} \\ 
+\frac 14\left( -\alpha \mid 0_10_2\rangle -\beta \mid 0_11_2\rangle -\gamma
\mid 1_10_2\rangle -\delta \mid 1_11_2\rangle \right) \Phi
_{35}^{-}\triangledown \Phi _{46}^{+} \\ 
\left( \alpha \mid 0_10_2\rangle -\beta \mid 0_11_2\rangle +\gamma \mid
1_10_2\rangle -\delta \mid 1_11_2\rangle \right) \Phi _{35}^{-}\triangledown
\Phi _{46}^{-}
\end{array}
\right\}
\end{equation}
We define eight $2\times 2$ unitary matrices by 
\begin{eqnarray}
U_{\Psi _{35}^{+}} &=&i\sigma _y,U_{\Psi _{35}^{-}}=-\sigma _x,U_{\Phi
_{35}^{+}}=\sigma _z,U_{\Phi _{35}^{-}}=-\sigma _0  \nonumber \\
U_{\Psi _{46}^{+}} &=&\sigma _x,U_{\Psi _{46}^{-}}=-i\sigma _y,U_{\Phi
_{46}^{+}}=\sigma _0,U_{\Phi _{46}^{-}}=-\sigma _z
\end{eqnarray}
where $\sigma _x=\left[ 
\begin{array}{ll}
& 1 \\ 
1 & 
\end{array}
\right] ,\sigma _y=\left[ 
\begin{array}{ll}
& -i \\ 
i & 
\end{array}
\right] ,\sigma _z=\left[ 
\begin{array}{ll}
1 &  \\ 
& -1
\end{array}
\right] ,\sigma _0=\left[ 
\begin{array}{ll}
1 &  \\ 
& 1
\end{array}
\right] $ are the Pauli matrices. Now $\mid \Psi _{123456}\rangle $ can be
written simply as 
\begin{eqnarray}
&\mid &\Psi _{123456}\rangle =\sum\nolimits_{K=\Psi _{35}^{+},\Psi
_{35}^{-},\Phi _{35}^{+},\Phi _{35}^{+}}\sum\nolimits_{L=\Psi _{46}^{+},\Psi
_{46}^{-},\Phi _{46}^{+},\Phi _{46}^{+}}\left( \frac 14\mid \varphi
_{K\triangledown L}^{*}\rangle \otimes \left( K\triangledown L\right) \right)
\nonumber \\
&\mid &\varphi _{K\triangledown L}^{*}\rangle =U_K\otimes U_L\left( \mid
\varphi _{12}\rangle \right)
\end{eqnarray}
where $\mid \varphi _{12}\rangle =\mid \varphi _{56}\left( 5\longrightarrow
1,6\longrightarrow 2\right) \rangle $. When Alice makes a Bell measurement
of particle pair $\left( 3,5\right) $, and a Bell measurement of particle
pair $\left( 4,6\right) $ respectively, the wave function will collapse to
one $\mid \varphi _{K\triangledown L}^{*}\rangle \otimes \left(
K\triangledown L\right) $ with probability $\frac 1{16}$ ($K\triangledown L$
is measured by Alice, simultaneously Bob obtain a corresponding state $\mid
\varphi _{K\triangledown L}^{*}\rangle ).$ When Alice informs Bob of her
measurement result (one $K\triangledown L)$ by a classical communication,
then Bob at once knows that the correct result should be 
\begin{equation}
\mid \varphi _{12}\rangle =\left( U_K\otimes U_L\right) ^{-1}\left( \mid
\varphi _{K\triangledown L}^{*}\rangle \right) =U_L^T\otimes U_K^T\left(
\mid \varphi _{K\triangledown L}^{*}\rangle \right)
\end{equation}
where $T\ $is the transposition. So, the bipartite qubit entanglement
teleportation has been completed. If we take other cross Bell basis as
channel, the steps are similar.

We see that in our method the order (13 and 24) of particles in cross Bell
basis is important, in fact, if we use the natural order (12 and 34), e.g. $%
\mid \Psi _{12}^{\pm }\rangle ,\mid \Phi _{34}^{\pm }\rangle ,\cdots ,$
etc., the product $\triangledown $ becomes common tensor product $\otimes $,
and we still choose a product of them as the quantum channel, then the
process in the above scheme will lead to inconveniency and difficulty.

{\it Discussion and conclusion. }If it is known that when $\alpha \gamma
-\beta \delta =0,$ then $\mid \varphi _{56}\rangle $ must be decomposed[5]
in form as $\mid \varphi _{56}\rangle =\left( a\mid 0_5\rangle +b\mid
1_5\rangle \right) \otimes \left( c\mid 0_6\rangle +d\mid 1_6\rangle \right) 
$, then obviously the above process, in fact, becomes two independent
teleportation of $\mid \varphi _5\rangle =a\mid 0_5\rangle +b\mid 1_5\rangle 
$ and $\mid \varphi _6\rangle =c\mid 0_6\rangle +d\mid 1_6\rangle $
respectively. In addition, for three states $\mid \Psi _{14}\rangle
=\sum\limits_{i,j=0,1}c_{ij}\mid i_1\rangle \mid j_4\rangle \in H_1\otimes
H_4,$ $\mid \Phi _{25}\rangle =\sum\limits_{r,s=0,1}d_{rs}\mid r_2\rangle
\mid s_5\rangle \in H_2\otimes H_5$ and $\mid \Omega _{36}\rangle
=\sum\limits_{x,y=0,1}e_{xy}\mid x_3\rangle \mid y_6\rangle \in H_3\otimes
H_6$ if we define the cross product 
\begin{equation}
\mid \Psi _{14}\rangle \triangledown \mid \Phi _{25}\rangle \triangledown
\mid \Omega _{36}\rangle =\sum_{i,r,j,s=0,1}c_{ij}d_{rs}e_{xy}\mid
i_1\rangle \mid r_2\rangle \mid x_3\rangle \mid j_4\rangle \mid s_5\rangle
\mid y_6\rangle \in \bigotimes\limits_{m=1}^6H_m
\end{equation}
and construct the cross Bell basis $\left\{ K\triangledown L\triangledown
M\right\} $ of $\bigotimes\limits_{m=1}^6H_m$, where $K,L,M=\mid \Psi
_{14}^{\pm }\rangle ,\mid \Phi _{14}^{\pm }\rangle ,$\ $L=\mid \Psi
_{25}^{\pm }\rangle ,\mid \Phi _{25}^{\pm }\rangle $,$\;M=\mid \Psi
_{36}^{\pm }\rangle ,\mid \Phi _{36}^{\pm }\rangle $, etc., then by a
similar way we can realize the teleporation of a unknown tripartite qubit
state. Obviously this method can be generalized to arbitrary dimensional
cases.

To sum up, by using cross Bell bases and Bell measurements we give a simple
scheme of arbitrary multipartite qubit states.

\end{document}